\documentclass[12pt]{article}
\usepackage[centertags]{amsmath}
\usepackage{amssymb}
\usepackage{graphicx,epsfig,graphics}

\begin{document}

\title{\bf Nonequilibrium phase transition in a mesoscoipic biochemical
system: From stochastic to nonlinear dynamics and beyond}

\author{Hao Ge$^1$\footnote{Email: gehao@fudan.edu.cn} \quad and \quad
Hong Qian$^{1,2}$\footnote{Email: qian@amath.washington.edu}\\[8pt]
$^1$School of Mathematical Sciences\\
and Center for Computational Systems Biology\\
Fudan University, Shanghai 200433, PRC\\
and\\
$^2$Department of Applied Mathematics\\
University of Washington, Seattle, WA 98195,
USA}

\maketitle{}

\vskip 0.5cm
\centerline{\bf\large Abstract}

    A rigorous mathematical framework for analyzing
the chemical master equation (CME) with bistability, based on
the theory of large deviation, is proposed.  Using
a simple phosphorylation-dephosphorylation cycle with
feedback as an example, we show that a nonequilibrium
steady-state (NESS) phase transition occurs in the
system which has all the characteristics of classic
equilibrium phase transition: Maxwell construction,
discontinuous fraction of phosphorylation as a function
of the kinase activity, and Lee-Yang's zero for the
generating function. The cusp in nonlinear bifurcation
theory matches the tricritical point of the phase
transition.  The mathematical analysis suggests three
distinct time scales, and related mathematical
descriptions, of (i) molecular signaling, (ii)
biochemical network dynamics, and (iii) cellular
evolution.  The (i) and (iii) are stochastic while
(ii) is deterministic.

\centerline{\rule{4in}{0.3mm}}

\vskip 0.5cm

    Microscopic, stochastic molecular fluctuations disappear in the
thermodynamic limit in which deterministic nonlinear behavior arises.
However, in the mesoscopic world of cellular biology, complex dynamics
with multiple time scales makes the meaning of {\em thermodynamic
limit} only relative of one level of description with respect to another.
More specifically, we shall show in the present paper that there are
three biologically significant time scales, with related levels of
mathematical description:
(i) stochastic molecular signaling,
(ii) deterministic biochemical network dynamics, and
finally (iii) stochastic (again!) cellular evolution.  In other words,
there is stochastic behavior beyond the deterministic dynamics; all
three levels are contained in a mesoscopic living system.
Current cellular molecular biology chiefly focuses on (i) while
increasingly interested in (ii); however, it is the (iii), we believe,
that is most relevant to major cellular biological issues such as
differentiation, apoptosis, and cancer immunoediting.

    Our conclusion is reached through a detailed mathematical
analysis of a simple cellular signaling module: a
phosphorylation-dephosphorylation cycle (PdPC)
with feedback \cite{ferrell_qian}.
We use the chemical master equation (CME) as the starting model.
In recent years CME has emerged as one of the physiochemical foundations
of cellular biochemistry \cite{CME_recent}.  The theory had
begun in 1940 and went through a major development in 1960s and
70s \cite{CME_history}.  In particular, the Brussels school
has used this theory as a mathematical bases of nonequilibrium
steady state (NESS), a term first proposed by Klein \cite{NESS}.
It is now widely accepted that both concepts of CME and NESS
are appropriate for studying isothermal, homeostatic cellular
biochemistry \cite{ener_diss}.  The mathematical theory of NESS is an
irreversible, but stationary stochastic processes, associated
with which the concepts of entropy production and stationary
distribution naturally arise \cite{lebowitz_jqq}.

    It is generally believed that the deterministic nonlinear dynamics,
derived from the CME in the limit of the reaction system volume
$V\rightarrow\infty$ according to Kurtz's theory \cite{kurtz}, is
the macroscopic counterpart of the chemical reaction system
\cite{CME_recent}. While this is certainly true, here we refine this
notion by studying the large-deviation properties of
$V\rightarrow\infty$, i.e., the thermodynamic limit.  We shall show
that in the case of a nonlinear dynamical system with multiple
dynamic attractors, there is a unique macroscopic thermodynamic
state; all the other macroscopic attractors are in fact metastable,
with an infinitesimal stationary probability $\propto e^{-\beta V}$
and exponential small exit rate $\propto e^{-\alpha V}$
$(\alpha,\beta >0)$.

    The mathematical theory of large deviation (LDT) \cite{LDT} is
the natural device for understanding the thermodynamic limit of
systems with mutlistability, i.e., phase transition(s)
\cite{larg_dev}. Our result based on the LDT rigorously establishes
the stochastic dynamics with bi(multi)-modal distribution as the
mesoscopic signature of a nonlinear dynamics with
bi(multi)-stability.

    We have recently re-examined the nonlinear bistability in
the context of biochemical signaling module \cite{MaxCons}.  In the
thermodynamic limit when $V$ tends infinity, there is a phase
transition associated with the conventional nonlinear dynamic
approach based on the Law of Mass Action, which is the macroscopic
limit, in some sense, of the CME \cite{kurtz}.  A Maxwell-type
construction is an integral part of a complete theory of the CME
\cite{MaxCons}.

    In equilibrium phase transition, Lee-Yang theorem for grand
canonical partition function is widely considered to be a deep and
elegant result \cite{lee-yang}.  We shall show,
non-differentiability of a function $c(\lambda)$ (the NESS
counterpart of the free energy function) is the origin of
multi-phase behavior, and it is because a zero of $G(\lambda)$ (the
NESS counterpart of a partition function) reaches the real axis.
Different attempts have been made to generalize the Lee-Yang theory
to NESS and to bimodalities in \cite{ness-lee-yang}.

{\bf Large deviation theory, Maxwell construction and first-order
phase transition in a NESS.}
We now consider the same biochemical signaling system in
\cite{ferrell_qian,MaxCons} in terms of a one-dimensional CME.
Let $p_V(n)$ as its stationary probability for $N_V$,
the random variable representing the activated kinase molecule
$X$; $V$ being the volume of the system.

According to the classic result of LDT \cite{LDT,larg_dev},
especially Sec. 4.5.2 in the text by Dembo and Zeitouni's text, it
concludes that if $\frac{N_V}{V}$ satisfies the LDT with a good
``rate function'' $\phi(x)$, i.e., $p_V(n)\sim e^{-V\phi(x)}$,
$x\geq 0$, then \\
(a) For each $\lambda$, the ``free energy function''
$c(\lambda)=\lim_{V\rightarrow\infty}\frac{1}{V}\log \langle
e^{\lambda N_V}\rangle$ exists, and it is finite and nondecreasing.
Moreover it satisfies
\begin{equation}
    c(\lambda)=\sup_{x\geq 0}\{\lambda x-\phi(x)\}.
\end{equation}
(b) If $\phi(x)$ is convex, then it is the Fenchel-Legendre
transform of $c(\lambda)$, namely,
\begin{equation}
    \phi(x)=c^*(x)\triangleq \sup_{\lambda\in \mathcal{R}}\{\lambda
x-c(\lambda)\}.
\end{equation}
(c) If $\phi(x)$ is not convex, then $c^*(x)$ is the
affine regularization of $\phi(x)$, i.e. $c^*(\cdot)\leq \phi(\cdot)$,
and for any convex rate function $f$ such that $f(\cdot)\leq
\phi(\cdot)$ implies $f(\cdot)\leq c^*(\cdot)$.

Consequently, we know that when $\phi(x)$ is bimodal with two local
minima, then they are at different heights if and only if the
$c(\lambda)$ is differentiable at $\lambda=0$ according to the
well-known G\"{a}rtner-Ellis theorem \cite{larg_dev}, and
$\frac{dc(0)}{d\lambda}$ is simply the position of the lower
minimum.  This implies that the Maxwell construction corresponds to
the function $c(\lambda)$ being non-analytic at $\lambda=0$.
Further, if the rate function $\phi(x)$ is analytic, then
$c(\lambda)$ is continuous and
\begin{equation}
    c(\lambda)=\sup_{\phi^{'}(x)=\lambda}\{\lambda x-\phi(x)\}.
\end{equation}

If a non-convex $\phi(x)$ has two local minima of $\phi(x)$ with
equal height, for sufficiently small $\lambda<0$, $c(\lambda)=\lambda
x-\phi(x)$ for the $x$ near the left minima satisfying
$\phi^{'}(x)=\lambda$; and when $\lambda>0$, also
$c(\lambda)=\lambda x-\phi(x)$ for the $x$ near the right minima
satisfying $\phi^{'}(x)=\lambda$. Therefore, the left and right
derivatives of the function $c(\lambda)$ at $\lambda=0$ both exist
but are equal to the left and right local minima respectively.

If $\phi(x)$ has two local minima $x_1$ and $x_2$ with different
heights, one can rewrite $\lambda x - \phi(x) = \lambda' x -
\psi(x)$ where $\psi(x)=\phi(x)-\lambda^*x$ such that $\psi(x)$ has
two minima with equal heights and $\lambda'=\lambda-\lambda^*$.
Hence, the nonanalytic point of $c(\lambda)$ moves to $\lambda^*$,
and also the left and right derivatives at $\lambda=\lambda^*$ both
exist and equal to the left and right local minima of $\psi(x)$
respectively. In other words, $\lambda^*$ is just the slope of the
tangent line of $\phi(x)$ with exactly two tangent points. More
generally, if the non-convex $\phi(x)$ has $k$ tangent lines with
more than one tangent points, then the function $c(\lambda)$ has $k$
non-analytical points, and vise versa. So it is a ``higher level''
of convexity! \cite{footnote1}

\begin{figure}[ht]
  \begin{center}
   \includegraphics[width=3in,angle=270]{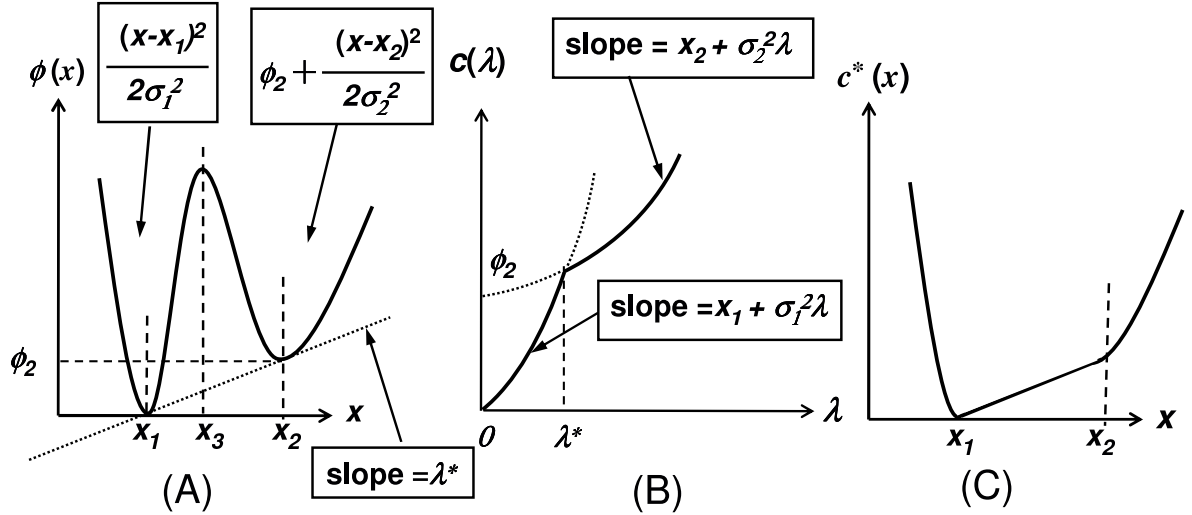}
\caption{Bimodal, non-convex $\phi(x)$ in (A) gives rise to
non-analyticity in the Fenchel-Legendre transform $c(\lambda)$ in
(B), the generating function of the $N_V$ in the thermodynamic
limit. In equilibiurm statistical mechanics, $c(\lambda)$ is the
free energy; hence the non-differentiability at $\lambda=\lambda^*$
indicates first-order phase transition. A quadratic function, with
curvature $\sigma^{-2}$ in (A) gives a quandratic function, with
curvature $\sigma^2$, in (B). $c^*(x)$ in (C) is the
Fenchel-Legendre transform of $c(\lambda)$, also known as the affine
regularization of $\phi(x)$.} \label{fig1}
  \end{center}
\end{figure}

    The above LDT results are summarized in Fig. \ref{fig1}.
Fig. \ref{fig1} shows that, as that in Lee-Yang's theory
\cite{lee-yang}, $c(\lambda)$ is continuous but non-differential at
$\lambda=\lambda^*$.

    Now let us consider another parameter $\theta$ of the system.
Let it be a bifurcation parameter in the nonlinear dynamics
according to the Law of Mass Action \cite{ferrell_qian}.  We
have shown in \cite{MaxCons} that the stable and unstable
fixed points of the nonlinear dynamics correspond precisely with
the minima and maxima of the $\phi(x)$, and bistability corresponds
to double-wells in $\phi(x)$, and bimodality of $-\phi(x)$.

\begin{figure}[ht]
  \begin{center}
   \includegraphics[width=3.5in,angle=270]{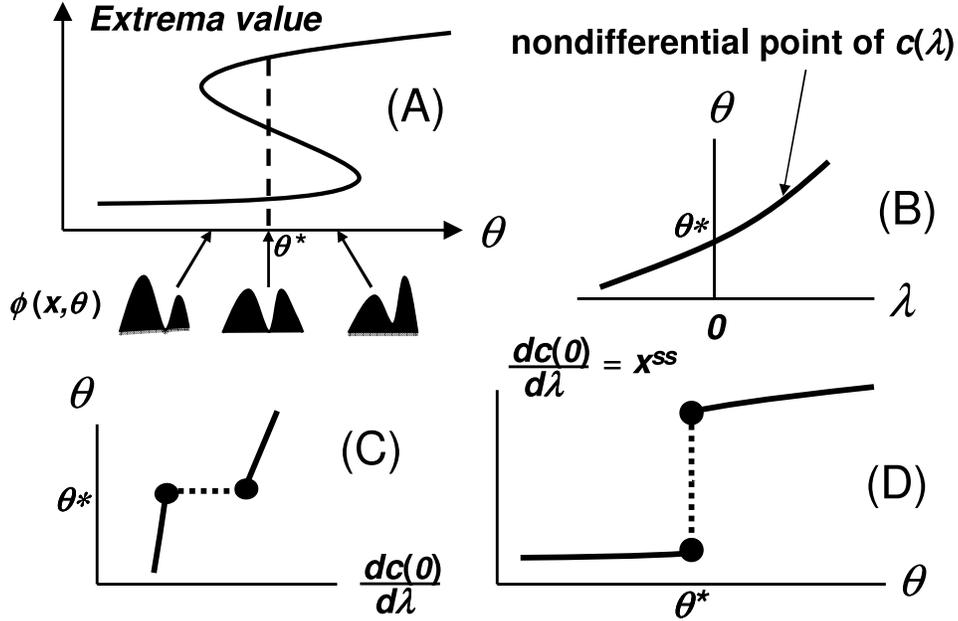}
   \caption{(A) The solid line represents the extrema of the
$\phi(x)$, which corresponds to the stable and unstable fixed points
of the nonlinear differential equation model.  $\theta$ is a
bifurcation parameter.  When $\theta=\theta^*$, the two wells of
$\phi(x)$ have equal height. (B) For each value of $\theta$, the
double-well $\phi(x)$ yields a non-analytical point
$\lambda^*(\theta)$.  This line crosses the $\lambda=0$ when
$\theta=\theta^*$.  (C) $\theta$ as a function of
$\frac{dc(0)}{d\lambda}$.  (D) $\frac{dc(0)}{d\lambda}$ is in fact
the position of the lower minima of $\phi(x)$, which is the mean
concentration of $X$ in the system. } \label{fig2}
  \end{center}
\end{figure}

    Here consider the function  $(1/V)\log \langle e^{\lambda N_V}
\rangle=c_V(\lambda,\theta)$. As $V$ tends to infinity, the limit
$c(\lambda,\theta)$ exists, and it is continuous and a
non-decreasing function of $\lambda$. Furthermore, there is a line
in the $(\lambda, \theta)$ plane at which the $c$ is
non-differentiable with respect to $\lambda$. The line passes
$(0,\theta^*)$ where $\theta^*$ is the critical value of Maxwell
construction with which the function $\phi(x)$ has two minima with
equal heights. Such a ``singularity line'' in the $\lambda-\theta$
space divides the space into upper left and lower right parts. They
represent two phases.  See Fig. \ref{fig2}.

In our theory, the derivative at $\lambda=0$ is particularly
meaningful: It is the mean concentration of molecules in the system
(property of the generating function). In the thermodynamic limit,
the mean and the highest peak position of $e^{-V\phi(x)}$ are the
same, the macroscopic value. Thus, we understand that the Maxwell
construction implies the mean concentration is not continuous.

{\bf Generalizing Lee-Yang's theory.}  In equilibrium phase
transition, according to \cite{lee-yang}, the non-analyticity in the
free energy function $c(\lambda)$ is due to a zero in the partition
function $G_V(\lambda)=\langle e^{\lambda N_V}\rangle$ approaching
the real axis from the complex plane of $\lambda$. Is the
non-analyticity in our $c(\lambda)$ also due to the zero of
$G(\lambda) = \lim_{V\rightarrow\infty}G_V(\lambda)$? This is indeed
the case.

Our probability distribution for $N_V$ has a finite support. So the
generating function is a finite order polynomial of z, (use $z =
e^{\lambda})$.  Then consider a region of the complex plane of $z$,
which contains a section of the $z$ axis. According to Theorem 2 in
\cite{lee-yang} which is a pure mathematical result, the zero
of the generating function must be ``pinched'' onto real z-axis at
the non-analytic point of the free energy function $c(\lambda)$ when
$V$ tends to infinity. Therefore, our theory generalizes the
Lee-Yang theory to nonequilibrium phase transition.

Several previous works have generalized the Lee-Yang theory in
nonequilibrium steady states \cite{ness-lee-yang} through specific
examples. It has been suggested that the bimodal distribution could
imply the Lee-Yang theory, but not vice versa.  This is consistent
with our result.

{\bf Cusp catastrophe and tri-critical point in
a PdPC with feedback.}  We consider the simple PdPC
with positive feedback which exhibits
nonlinear bistability \cite{ferrell_qian}:
\begin{equation}
E+K^* \rightleftharpoons E^*+K^*,
    \ \
K+2E^* \rightleftharpoons K^*,
         \ \
E^*+P \rightleftharpoons E+P,
\end{equation}
in which $K$ and $K^*$ are inactive and active forms of a kinase,
$P$ is a phosphatase. $E^*$ is the phosphorylated $E$, a signaling
molecule.  Usually $E^*$ is functionally active, i.e., ``turned-on''.
Following the previous treatment \cite{ferrell_qian,MaxCons},
we assume the reversible binding
$K+2E^*\rightleftharpoons K^*$ is rapid.  Hence, the
dynamics of the fraction of phosphorylated $E$, $x$, satisfies
\begin{equation}
    \frac{dx}{dt} = \theta x^2\left[(1-x)- \epsilon x\right]+
            \left[\mu(1-x)-x\right] = r(x;\theta,\epsilon),
\label{pdpcwfb}
\end{equation}
in which the three parameters $\theta$ represents the ratio of the
activity of the kinase to that of the phosphatase; $\epsilon$
represents the ADP to ATP concentration ratio, and $\mu$ represents
the strength of phosphorolysis. $-k_BT\ln(\mu\epsilon)=\Delta G$
represents the ATP hydrolysis energy.  In a living cell, both $\mu$
and $\epsilon$ are small; hence $\gamma=\frac{1}{\mu\epsilon} \gg
1$.

    For large system's volume $V$, the CME gives the
stationary probability $p^{ness}(x)\propto e^{-V\phi(x)}$,
where the LDT rate function \cite{MaxCons}
\begin{equation}
    \phi(x) = \ln (1-x)-x\ln\left[\frac{(1-x)(\theta x^2+\mu)}
        {x(\theta\epsilon x^2+1)}\right]
        +2\sqrt{\frac{\theta}{\mu}}\arctan
    \left(\sqrt{\frac{\theta}{\mu}}x\right)
        -\frac{2}{\sqrt{\theta\epsilon}} \arctan {\sqrt{\theta\epsilon}x}.
\end{equation}
One can easily check that
\begin{equation}
    \frac{d\phi(x)}{dx} = -\ln\frac{(1-x)(\theta x^2+\mu)}
            {x(\theta\epsilon x^2+1)},
\end{equation}
and the extrema match exactly with the roots of $r(x;\theta,\epsilon)=0$.

    The Eq. (\ref{pdpcwfb}) exhibits
saddle-node bifurcations and cusp catastrophe.  One obtains the
parameter region for the bistability from simultaneously solving
$r(z)=0$ and $\frac{dr(z)}{dz}=0$:
\begin{equation}
    \theta z^2\left[1-(1+\epsilon)z\right]+
            \left[\mu-(1+\mu)z\right] = 0, \ \ \
    \theta \left[2z-3(1+\epsilon)z^2\right]
        -(1+\mu) = 0.
\end{equation}
The two equations give the boundary of the region of bistability in
$(\theta,\epsilon)$ space  (in terms of $z$ as a parametric
curve):
\begin{equation}
    \theta = \frac{2(1+\mu)}{z}-\frac{3\mu}{z^2}, \ \ \
         \epsilon=\frac{2\mu-(\mu+1)z}{3\mu z-2(\mu+1)z^2}-1.
\end{equation}

\begin{figure}[ht]
  \begin{center}
   \includegraphics[width=4in,angle=270]{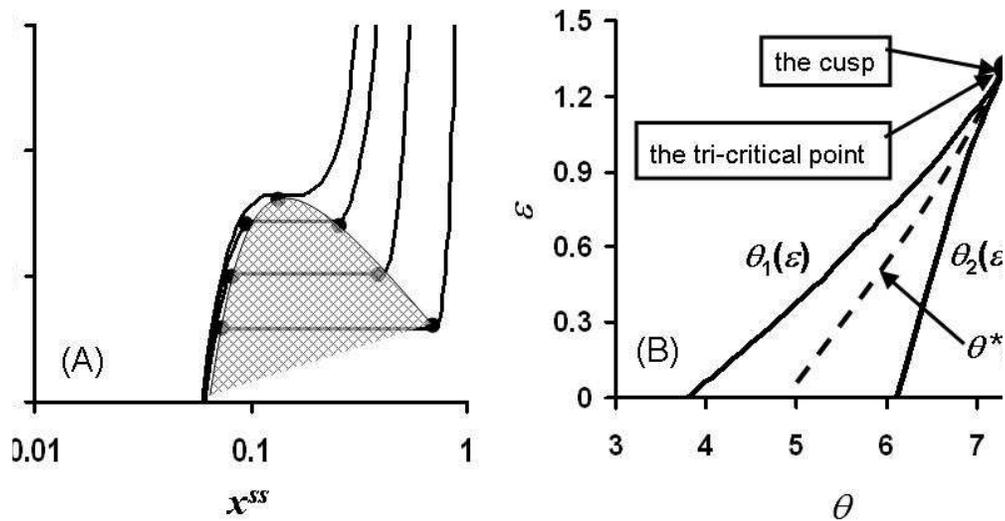}
   \caption{(A) Steady state $x^{ss}$ as functions of $\theta$ according
to Eq. (\ref{pdpcwfb}), together with Maxwell constructions under the
shaded region.  The parameter used $\mu=0.05$, with $\epsilon=1.3,1,0.5,0.005$
from top to bottom. (B) The solid lines represent the saddle-node critical
points, i.e., the filled circles in (A).  They meet at a cusp.
The dashed line represents the
critical value at which the Maxwell construction is performed.  The
dashed line terminates at the cusp.
} \label{fig3}
  \end{center}
\end{figure}

    Fig. \ref{fig3}A shows the steady states of Eq. (\ref{pdpcwfb}), $x^{ss}$,
as a function of $\theta$ with various $\epsilon$.  We see for the
range of $\epsilon\le 1.33$ the system has three fixed points, i.e.,
bistability.  After introducing the Maxwell construction for each
and every curve $x^{ss}(\theta)$, we obtain a set of monotonic
$x^{ss}(\theta)$.  This corresponds to the ``PV-isotherm'' in the
van der Waals theory of phase transition.

    According to the cusp catastrope theory \cite{Catastrophes},
there is a region in the $\epsilon-\theta$ plane with three fixed
points.  The boundary of the region is where the system has exactly
two fixed points, i.e., where bifurcation occurs:
$\theta_1(\epsilon)$ and $\theta_2(\epsilon)$. One of the most
important features of this region is that it has a cusp, at
$\theta_{cusp}=\frac{(1+\mu)^2}{3\mu}$,
$\epsilon_{cusp}=\frac{1-8\mu}{9\mu}$, when
$z_{cusp}=\frac{3\mu}{1+\mu}$, as shown in Fig. \ref{fig3}B.

    For a given $\epsilon$, the critical $\theta^*$ at
which the Maxwell construction is performed satisfies
$\theta_1\le \theta^*\le \theta_2$.   Thus, the critical line
$\theta^*(\epsilon)$ abruptly terminates at the cusp.
In equilibrirum phase transition, the cusp is also known as
tri-critical point \cite{Gaite}.

    We also note that bistability implies that the $x^{ss}$ as
function of the $\theta$, or $\epsilon$, is not monotonic (it is
S-shaped).  However, after the Maxwell construction the resulting
$x^{ss}(\theta)$ is monotonic in ``true'' thermodynamic limit! It is
precisely the same situation as the PV isotherm in gas-liquid phase
transition. The word ``true'' means one has to wait sufficiently
long to allow the jumps back and forth between attractors. The
biological significance of monotonicity remained to be elucidated.

{\bf Discussion.}  The present paper shows that many classical
concepts from equilibrium phase transition can be applied to
bifurcation problem in nonlinear chemical dynamics that
has a mesoscopic stochastic
underline in terms of the CME. From the CME point of view,
the LDT treatment we present is a small but significant step
beyond the Kurtz theory towards the macroscopic nonlinear
dynamics.  Analyzing the CME is a much more challenging
problem than analyzing the partition function since the
former offers a dynamic theory.

    The celebrated Maxwell construction is a natural consequence
of the general theory we propose, and the well-known Lee-Yang theorem
is in fact a special case of it.  More importantly, the general theory
is applicable to driven systems with nonequilibrium steady state.

    On the mathematical side, the general theory provides a
framework to study nonlinear bifurcations in terms of mathematical
non-analyticity of a certain function, a vision long being hold by
some investigators \cite{Zeeman}. The large deviation function
$\phi(x)$ can be in fact considered as some type of stochastic
{\em landscape} (potential, Lyapunov function in a not rigorous sense)
for systems without gradient, nor detailed balance \cite{ref_on_potential}.

    While the CME as a fundamental theory of studying cellular
biochemistry remains to be validated experimentally, it is certainly
an acceptable mathematical model for studying mesoscopic complexity
and emergent organization, as called by Laughlin et al.
\cite{Laughlin}. Chemical reactions are marvellous systems for
understanding complexity. The present work shows that while Kurtz's
theorem is correct, the real limit of V tends infinity is not the
solution to the law of mass action, but rather requires a LDT
treatment.

  The existence of ``nice'' $\phi(x)$ in the asymptotic form of
$e^{-V\phi(x)}$
is not always true for the CME; note that there are chaotic behavior
as well involved. If one considers a CME whose corresponding
ODE is a 3-dimensional chaotic dynamics with a strange attractor,
what will be the stationary
distribution in the limit of $V\rightarrow\infty$? This problem
has been discussed in the past \cite{hugang}.
The general feeling is that $\phi(x)$ is not smooth itself. So one does
not have a ``nice'' $\phi(x)$! For a very ``rugged $\phi(x)$'', we
believe that its Fenchel-Legendre transform $c(\lambda)$ might be a
very powerful way to ``find the key feature'' of the $\phi(x)$. The
number of non-differentiable point is definitely much smaller than
the number of peaks!

{\em Beyond deterministic dynamics.}
It is generally believed that when a
system's size increases, the stochastic behavior at a mesoscopic
level averaged out, and a deterministic behavior emerges.   However,
our present analysis clearly show that the emerging deterministic
behavior in the CME is a metastable system's dynamics.  Beyond that
time scale, another ``macroscopic'' {\em stochastic} behaviour exists!
This multi-attractor stochastic system is a true emerging phenomenon
that one can not naively expect from the deterministic dynamics (e.g.,
based on the relative area of the attractive basins) without detailed
stochastic mechanistic modeling. The Maxwell construction is
the consequence of the steady state on this
``beyond-deterministic-infinite'' time scale.

There are three time scales in this mathematical hierarchy of cellular
dynamics:  A molecular signaling time scale
(i.e., the rate constant for molecular interactions), a biochemical
network time scale (i.e., the deterministic relaxation times to attractors),
and a cellular evolutionary time scale). We believe it is at the last level
of stochastic dynamics that is most relevant to major cellular biological
issues such as differentiation, apoptosis, and cancer immunoediting.

\begin{figure}[ht]
  \begin{center}
  \includegraphics[width=4in,angle=270]{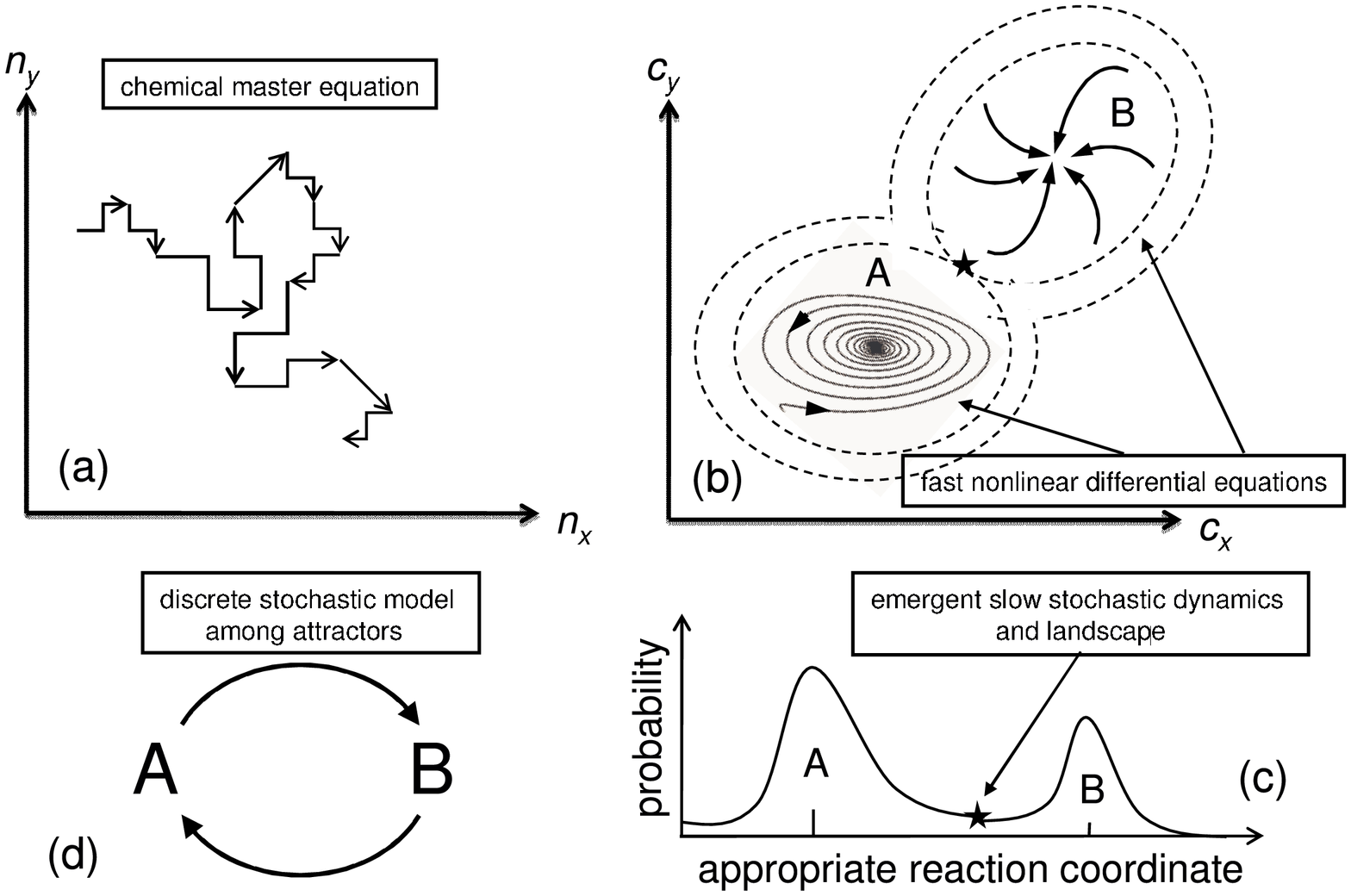}
   \caption{Schematics showing the mathematical hierarchy of
cellular dynamics based on the chemical master equation (CME)
approach.  (a) stochastic dynamics based on the Gillespie
algorithm; (b) deterministic dynamics tending to attractors;
(c) probabilistic distributions for the two attractors;
(d) stochastic dynamics among the attractors.  (a), (b) and
(c) represent stochastic molecular signaling,
deterministic biochemical dynamics, and
stochastic cellular evolution, respectively.
}
\label{fig4}
\end{center}
\end{figure}

\section{Acknowledgement}

HQ thanks Prof. R.M. O'Malley for many helpful discussions.
While it is obvious that the problem presented here is
intimately related to Laplace integrals and turning-point
problems (the star in Fig. \ref{fig4}C), a more comprehensive
application of the singular purturbation approach remains to
be worked out.


\begin{thebibliography}{99}

\bibitem{ferrell_qian}
J.E. Ferrell and W. Xiong, Chaos, {\bf 11}, 227 (2001);
H. Qian and T.C. Reluga, Phys. Rev. Lett. {\bf 94} 028101 (2005);
H. Qian, Ann. Rev. Phys. Chem. {\bf 58}, 113 (2007).


\bibitem{CME_recent}
M.S. Samoilov and A.P. Arkin, Nat. Biotechnol. {\bf 24}, 1235 (2006);
D.T. Gillespie, Ann. Rev. Phys. Chem. {\bf 58}, 35 (2007);
D.A. Beard and H. Qian, {\em Chemical Biophysics: Quantitative Analysis
of Cellular Systems}  (Cambridge Univ. Press, London 2008).

\bibitem{CME_history}
M. Delbr\"{u}ck, J. Chem. Phys. {\bf 8}, 120 (1940);
D.A. McQuarrie, D.A. J. Appl. Prob. {\bf 4}, 413 (1967);
G. Nicolis and I. Prigogine, {\em Self-Organization in Nonequilibrium
Systems: From Dissipative Structures to Order Through Fluctuations}
(Wiley, New York, 1977).

\bibitem{NESS}
P. Gaspard, J. Chem. Phys. {\bf 120}, 8898 (2004);
M. Lax, Rev. Mod. Phys. {\bf 32}, 25 (1960);
M.J. Klein, Phys. Rev. {\bf 97}, 1446 (1955).

\bibitem{ener_diss}
R. Phillips, J. Kondev, and J. Theriot,
{\em Physical Biology of the Cell} (Garland Science, New York, 2008);
E. Karsenti, Nature Rev. Mol. Cell Biol. {\bf 9}, 255 (2008);
H. Qian, J. Phys. Chem. B {\bf 110}, 15063 (2006).

\bibitem{lebowitz_jqq}
D.-Q. Jiang, M. Qian, M.-P. and Qian,  {\em Mathematical Theory of
Nonequilibrium Steady States - On the Frontier of Probability and Dynamical
Systems.}  (Lect. Notes Math. Vol. 1833, Springer-Verlag, Berlin, 2004);
P.G. Bergmann and J.L. Lebowitz, Phys. Rev. {\bf 99}, 578 (1955).

\bibitem{kurtz}
T.G. Kurtz, J. Chem. Phys. {\bf 57}, 2976 (1972).

\bibitem{LDT}
A. Dembo and O. Zeitouni, {\em Large Deviations Techniques and
Applications}, 2nd Ed. (Springer-Verlag, New York, 1998)

\bibitem{larg_dev}
H. Touchette, arXiv:0804.0327 (2008);
R.S. Ellis, {\em Entropy, Large Deviations, and Statistical Mechanics.}
(Springer-Verlag, New York, 1985).

\bibitem{MaxCons}
H. Ge and H. Qian, arXiv:0904.2056 (2009);
M. Vellela and H. Qian, J. Roy. Soc. Interf. In the press (2009).

\bibitem{footnote1}
We observe that if $\phi(x)$ has three minima, and the middle one is
the highest, it seems that the $c(\lambda)$ will be insensitive to
the middle minimum!  It only depends on the two low ones! However,
if the middle one is the lowest, then $c(\lambda)$ will have two
non-analytical points.

\bibitem{lee-yang}
C.N. Yang and T.D. Lee, Phys. Rev. Lett. {\bf 87}, 404 (1952); T.D.
Lee and C.N. Yang, Phys. Rev. Lett. {\bf 87}, 410 (1952); B.H. Zimm,
J. Chem. Phys. {\bf  19}, 1019 (1951).

\bibitem{ness-lee-yang}
R.A. Blythe and M.R. Evans, Phys. Rev. Lett. {\bf 89}, 080601
(2002); Ph. Chomaz, and F. Gulminelli, Physica A, {\bf 330}, 451
(2003); H. Touchette, Physica A, {\bf 359}, 375 (2005).


\bibitem{Catastrophes}
R. Thom, {\em Stabilit$\acute{e}$ Structurelle et Morphogenese}
(Intereditions, Paris, 1977); C. Zeeman, in {\em Structural
stability, The Theory of Catastrophies, and Applications in the
Sciences,} Vol. 525 of {\em Lecture Notes in Mathematics,} edited by
A. Dold and B. Eckmann (Springer, Berlin, 1976), p. 328

\bibitem{Gaite}
J.A. Gaite, Phys. Rev. A {\bf 41}, 5320 (1990).

\bibitem{Zeeman}
E.C. Zeeman, Nonlinearity {\bf 1}, 115--155 (1988).

\bibitem{ref_on_potential}
R. Graham and H. Haken,  Zeit. Physik A  {\bf 243},  289--302 (1971);
G. Nicolis and R. Lefever, Phys. Lett. A. {\bf 62},  469--471 (1977);
J. Ross, K.L.C. Hunt and M.O.Vlad, J. Phys. Chem. A. {\bf 106}, 10951--10960 (2002);
P. Ao, J. Phys. A. Math. Gen. {\bf 37}, L25--L30 (2004).


\bibitem{Laughlin}
R.B. Laughlin, D. Pines, J. Schmalian, B.P.
Stojkovi and P.G. Wolynes,
Proc. Natl. Acad. Sci. U.S.A.  {\bf 97}, 32 (2000).

\bibitem{hugang}
G. Hu, Zeit. Physik B {\bf 65}, 103--106 (1986);
R. Graham and T. T\'{e}l, Phys. Rev. Lett. {\bf 52}, 9--12 (1984).

\end{thebibliography}
\end{document}